\def\year{2020}\relax
\documentclass[letterpaper]{article} 
\usepackage{aaai20}  
\usepackage{times}  
\usepackage{helvet} 
\usepackage{courier}  
\usepackage[hyphens]{url}  
\usepackage{graphicx} 
\usepackage{graphicx}  

\usepackage{enumitem}
\usepackage{multirow}
\usepackage{hhline}
\usepackage{booktabs}
\usepackage{subcaption}
\usepackage{stackengine}

\usepackage{color}
\usepackage{todonotes}
\usepackage{amsmath,amssymb}

\colorlet{ColorforSina}{green!5!orange!95!}

\title{Machine Learning Explanations to Prevent Overtrust in Fake News Detection}

\author{Sina Mohseni\textsuperscript{\rm 1},
Fan Yang\textsuperscript{\rm 1},
Shiva Pentyala\textsuperscript{\rm 1},
Mengnan Du\textsuperscript{\rm 1},
Yi Liu\textsuperscript{\rm 1},
Nic Lupfer\textsuperscript{\rm 1},
\\ \Large \textbf{
Xia Hu\textsuperscript{\rm 1},
Shuiwang Ji\textsuperscript{\rm 1},
Eric Ragan\textsuperscript{\rm 2}} \\
\textsuperscript{\rm 1} Texas A\&M University, College Station, TX \\
\textsuperscript{\rm 2} University of Florida, Gainesville, FL \\
\{sina.mohseni,nacoyang,pk123,yiliu,nlupfer,xiahu,sji\}@tamu.edu, eragan@ufl.edu
}

\begin{document}

\maketitle

\begin{abstract}

Combating fake news and misinformation propagation is a challenging task in the post-truth era. 
News feed and search algorithms could potentially lead to unintentional large-scale propagation of false and fabricated information with users being exposed to algorithmically selected false content.
Our research investigates the effects of an \textit{Explainable AI} assistant embedded in news review platforms for combating the propagation of fake news. 
We design a news reviewing and sharing interface, create a dataset of news stories, and train four interpretable fake news detection algorithms to study the effects of algorithmic transparency on end-users.
We present evaluation results and analysis from multiple controlled crowdsourced studies. 
For a deeper understanding of \textit{Explainable AI} systems, we discuss interactions between user engagement, mental model, trust, and performance measures in the process of explaining. 
The study results indicate that explanations helped participants to build appropriate mental models of the intelligent assistants in different conditions and adjust their trust accordingly for model limitations.



\end{abstract}

\section{Introduction}

\noindent Artificial intelligence (AI) algorithms are used in a variety of online applications from product recommendation and targeted advertisement to loan and insurance rate prediction. 
However, as AI-based decision-making is directly affecting people's lives, the accountability and fairness of advanced AI algorithms are under question. 
In recent years, the need for algorithmic transparency is gaining more attention to enable accountable AI-based decision-making systems. 
To this end, \textit{Explainable AI} (XAI) techniques~\cite{gunning2017explainable} have been introduced to annex transparency into black-box machine-learning algorithms. 
Interpretability can help users to build a mental model of how algorithms work and build appropriate trust in intelligent systems Rader et al. ~\shortcite{rader2018explanations}. 

In the social media domain, news feed and search algorithms function similar to decision-making algorithms, as users are exposed to algorithmically selected content~\cite{trielli2019search}.
Blindly trusting algorithmically-curated news could potentially lead to unintentional large-scale propagation of false and fabricated information with users being exposed to false content and its re-sharing through social media. 
Human review of news and data mining techniques for fake-news detection and debunking are commonly being practiced as primary approaches to reduce fake news in social media. 
However, reviewing the life cycle of news in social media reveals opportunities to combat the propagation of fake news within news-feed platforms~\cite{mohseni2019open}. 
For example, AI-based news review assistant tools can be embedded in news feed platforms and have the potential to benefit users by providing direct suggestions related to news credibility rather than automatic organizational news debunking.

Our research objective is to study the effects of transparency for fake news detection through an XAI assistant for news review applications and social media. 
We investigate whether the interpretability of the fake news detector algorithm could enhance users overall experience and result in increased credibility of user-shared news. 
We also aim to examine whether model explanations can help users to avoid overtrusting the fake news detector when explanations are nonsensical to users.
We formulate our goals into the following research questions:  
 
\begin{itemize}
\vspace{-0.04cm}\item RQ1: Do AI and XAI assistants help end-users share more credible news? 
\vspace{-0.04cm}\item RQ2: How do AI explanations affect users' mental models of intelligent assistants? 
\vspace{-0.04cm}\item RQ3: How do AI explanations affect end-user trust and reliance in intelligent assistants? 
\end{itemize}

To address these research questions, we designed a news reviewing and sharing interface with a built-in interpretable fake news detector (AI and XAI assistants) for end-users and run a series of evaluation experiments. %
The study results indicate the complexity of the fake news detection problem and limitations of current model interpretability techniques for this task. 
Though the addition of explanations to our system did not improve user task performance, we observed that explanations helped participants' to build appropriate mental models of the intelligent assistants in different conditions and adjust their trust accordingly for the model logic.


\section{Related Work}

Machine learning algorithms are heavily used in online platforms and social media to analyze user data for improving user experience and increasing corporate profit. 
However, the lack of transparency can raise data privacy and model trustworthiness concerns in critical domains, and hence potentially decreases user trust and confidence in the long run~\cite{mohseni2019survey}. In this regard, researchers study the communication of algorithmic processes in various domains such as online advertising~\cite{eslami2018Communicating}, social media feeds~\cite{eslami2015always}, and personalized news search engines~\cite{ter2017news}. 
In this section, we briefly review machine learning and human-computer interaction papers related to the explainable news feed and fake news detection systems.

\subsection{Interpretable Fake News Detection}

Reviewing machine learning techniques for misinformation detection shows diverse directions like style-based approaches to detect deception-oriented writing~\cite{rubin2015truth} and hyper-partisan content~\cite{potthast2017stylometric}. 
To incorporate more data for representation learning, Shu et al.~\shortcite{shu2017exploiting} included news publisher information, user stance, and user engagement together in their Tri-Relationship fake news detection framework. 
In other work, Popat et al. \shortcite{popat2018declare} used the Google search engine to directly collect similar instances from the web with a sentence similarity as the measure.  
Their technique leverages external news articles to gather evidence from online sources. 

Multiple research efforts have shown that algorithm interpretability could potentially improve user attitudes toward algorithms. 
Du et al.~\shortcite{du2019techniques} categorized interpretation methods to explain predictions of natural language processing (NLP) models into four categories of back-propagation based, perturbation based, local approximation based, and decomposition based techniques. 
Back-propagation based methods calculate gradients or variants of gradients of a model prediction with respect to each  input~\cite{hechtlinger2016interpretation}. 
Gradient values for each word input can represent its contribution to the model prediction.
Alternatively, in perturbation based techniques, the occlusion of input text can cause changes in the model prediction resulting to estimate word contributions~\cite{li2016understanding}. 
Different from the other two techniques, decomposition-based methods can model the data flow process by calculating the additive contribution of each input word to final prediction~\cite{murdoch2018beyond}. 
Lastly, local approximation-based methods can explain a complex model's predictions by approximating its behavior around an input instance~\cite{ribeiro2016should}.

Recent interpretable fake news systems have shown advantages of interpretability for fake news detection, including helping end-users to find model weaknesses so that users can build an appropriate level of trust in model predictions.
For instance, XFake detector in~\cite{yang2019XFake} uses a tree-based model visualization to explain the overall decision paths for input news instances. 
In another paper, Shu et al.~\shortcite{shu2019defend} provide feature-based explanations for important user comments from relevant news articles for user interpretability on fake news detection. 
However, While these models achieve moderate performance (i.e., below 80\% detection accuracy) in detecting fake news, it remains uncertain that how would model explanations help end-users in detecting fake news. 
In this paper, we designed a news review environment with built-in explainable to study the effects of algorithmic transparency on human-AI collaboration for fake news detection.

\subsection{Explainability for Users}

Multiple studies on transparent AI explore design choices to build accurate mental model of algorithms and adjust end-users trust in AI systems. 
For instance, Kocielnik et al.~\shortcite{Kocielnik2019will} investigate accuracy indicator, example-based explanation, and user control as design choices to improve human-AI collaboration. 
Their findings indicate that users' perception of control had a significant positive effect on user trust. 
In the evaluation of XAI interfaces, Poursabzi et al.~\shortcite{poursabzi2018manipulating} present a comprehensive evaluation study for users' mental model (via user prediction task) and trust (via user agreement with AI) in interpretable models. 
Their results indicate the positive effect of interpretability on participants' mental model, however, they did not observe improvement on user trust. 
On the other hand, Papenmeier et al.~\shortcite{papenmeier2019model} show users could potentially lose trust in AI when exposed to low fidelity explanations.

To gain insight into whether news recommendation algorithms should be transparent about their decisions, Hoeve et al.~\shortcite{ter2017news} run a survey and learn that a large majority of respondents prefer explanations. 
However, in a follow up A/B testing, they find participants are not opening (via click count) model explanations.
This could be due to the low urgency of explanations in news recommendation and/or their study news test set.
In human studies for AI-based news fact-checking, Horne et al.~\shortcite{horne2019rating} run an experimental human subject study and find that feature-based explanations in AI assistant significantly improve users perception of news bias. 
Their measured effect size was much larger for participants who were frequent news readers and those familiar with politics, though. 
In another paper, Nguyen et al.~\shortcite{nguyen2018believe} present design and evaluation of a mixed-initiative fact-checking system to blend human knowledge with machine learning algorithms.
They also conclude that transparency and interactivity significantly affect users' ability to predict the veracity of given claims.
To continue this line of research, we investigate how different types of model explanations affect the credibility of news shared by users in a social media like scenario.
We also measure a wider range of explicit and implicit user feedback to study interactions among key XAI design goals in the explaining process.


\section{System Design}

To serve our research goals for studying the role of interpretable models in fake news detection, we designed an interface for users to review news stories and articles as well as interact with a fake news detection assistant and its explanations. 
Our system's targeted users are the general public who read daily news and are not AI experts nor news analysts. 
We started with identifying candidates for useful and impactful explanations for fake news detection such as keyword attention, supporting evidence, and source credibility based on machine learning research on misinformation detection and human-computer interaction research on news feed systems~\cite{mohseni2019open}. 
We followed Mohseni et al.'s~\shortcite{mohseni2019survey} design framework for XAI systems, which involved design iterations with series of pilot user testing for both model explanations and news interface.

\subsection{Explainable Interface}

We designed an interface for users to review a queue of news stories, share true news for other users, and report fake news stories. 
Our interface design process started with multiple interface sketches that suit the news reviewing task. 
We aimed to design a simple interface with useful explanations for fake news detection. 
We tested mock-up implementations from the top design choices with a small number of participants.
After reviewing feedback from user observations and interviews, we selected the most comprehensible and conclusive design for our human-subject experiments.

\begin{figure}
\captionsetup[subfigure]{justification=centering}
\centering
       \begin{subfigure}[t]{0.99\columnwidth}
        \stackunder[5pt]{\frame{\includegraphics[width=1.0\columnwidth]{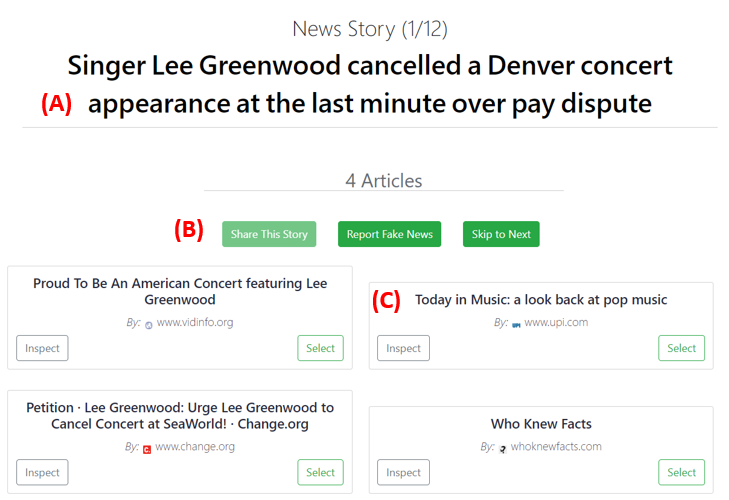}}}{Interface in Baseline condition.}
        \hfill
        \\
    \end{subfigure}
      \hfill
      \hfill
      \begin{subfigure}[t]{.99\columnwidth}
        \stackunder[5pt]{\frame{\includegraphics[width=1.0\columnwidth]{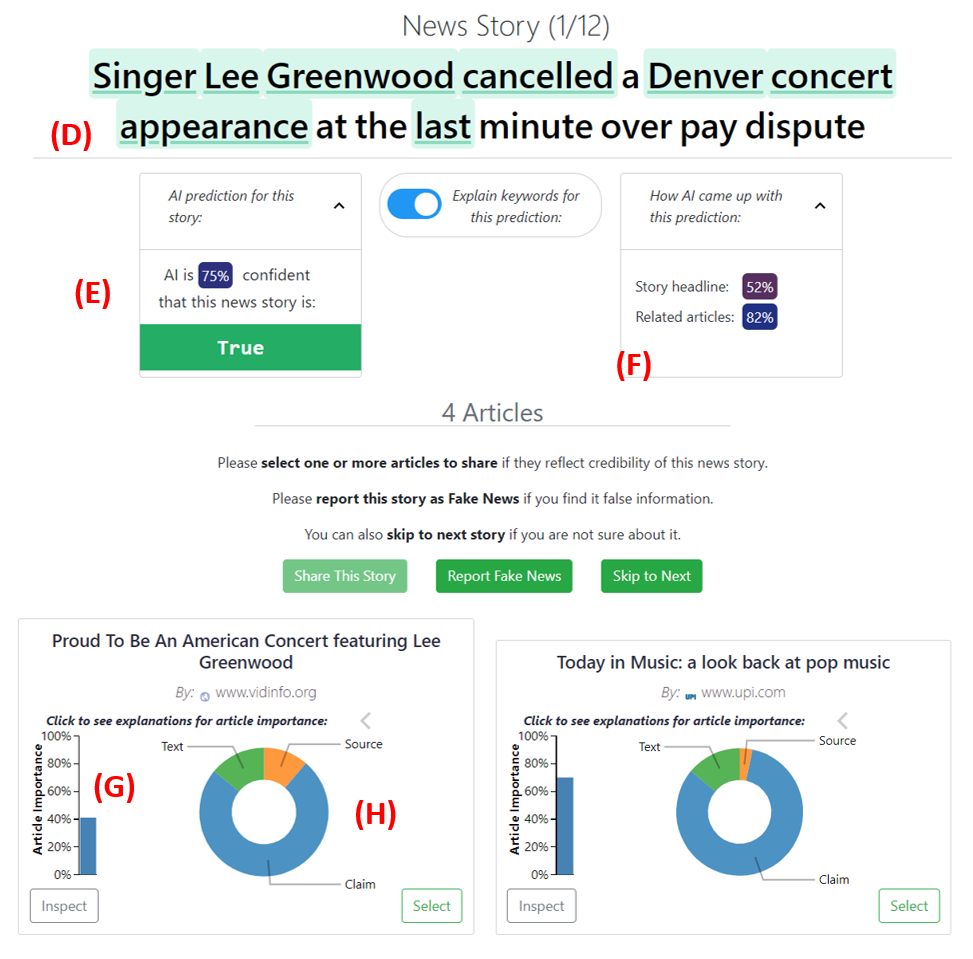}}}{Interface in XAI-all condition.}
    \end{subfigure}
  \caption{
  Our news review interface with AI and XAI assistants. 
  Interface components are removed for different study conditions.
  \textbf{Top:} Baseline interface without AI assistant. 
  (A) news headline.
  (B) user selecting to share, report, or skip the news story. 
  (C) a list of related news articles for the headline. 
  The user can open articles to inspect details. 
  \textbf{Bottom:} Interface with clickable model prediction and different feature attribution explanations. 
  (D) a heatmap of word level feature attribution explanation for news headline. User can see the attribution score values in tooltips when hovering mouse over the keywords.
  (E) fake news prediction and confidence. 
  (F) confidence for the headline and article separately. 
  (G) a bar chart for each article attribution score in comparison to other related articles. 
  Bar charts show lower values when the articles are less related to the headline or less significant for model prediction.  
  (H) A donut chart for each news article for source attribution scores compared to headline (Claim) and article (Text) content.
  }
  \label{fig:interface}
\end{figure}

Figure~\ref{fig:interface}-Top shows our baseline interface that enables the participants' news review task.
The interface shows a news headline for a news story on the top (Figure~\ref{fig:interface}-A) followed by a list of related articles below (Figure~\ref{fig:interface}-C). 
The related articles provide context and article sources for the news headline, and they can help the user to understand contributing information and factors for model prediction. 
The system allows users to open and read the related articles, but for our study, it was not required for sharing the news headline. 
The system was designed to allow users to review news stories one-by-one and decide if 1) the story is true to be shared with other users, or 2) it is fake news to be reported, or 3) they want to skip to the next story due to their unfamiliarity with the topic or lack of confidence (see Figure~\ref{fig:interface}-C).

Figure~\ref{fig:interface}-Bottom shows our interface with the XAI assistant. 
The fake news detection assistant is embedded in our interface which provides the model prediction (with or without explanation) about the news storys' credibility. 
Both the AI assistant prediction and its explanations are in the form of on-demand recommendations for the user, which are collapsible on user click. 
Our XAI assistant design provides \textit{why-type} explanations (see~\cite{mohseni2019survey} for discussion of explanation types) from the ensemble of fake news detectors. 
These explanations describe the attribution of different news features (i.e., news headline, article text, and article source) for each news veracity prediction. 
These different explanations are presented to the user with the following visual elements: 
(1) A heatmap of keyword attribution score that explains how the XAI assistant learned word-level features in the news headline (Figure~\ref{fig:interface}-D) and its related articles (in the news article page).  
(2) A single bar chart for each related article (Figure~\ref{fig:interface}-G) explaining each related article's attribution score in comparison to other articles for the model prediction. 
(3) A pie chart to present attribution score for the articles' source in comparison to the articles' content attribution and news headline attribution. 
(4) A list of top-3 important sentences for the article is shown when reviewing news articles to explain sentence-level feature learning of the models. 

The core design rationale for the four explanations was to embed model attribution explanations using visual elements for each news feature. 
Each visual explanation element was tested with users and refined through iterative design. 

\subsection{Fake News Data}
Training data for our models come from two sources: a) news story headlines and labels from Snopes (www.snopes.com) and b) related articles crawled from Google search results (top 16). 
The related articles were collected for each Snope news headline separately and labeled the same as their respective Snopes news story statements with noisy label assumption for the purpose of model training. 
The training data includes 4638 news story headlines with an average length of 15 words and 30599 related articles with an average length of 1012 words. 
We used 80\% of data for model training, 10\% data for validation, and 10\% data for testing. 
We took news samples and model predictions from our test set to feed the interface for human-subject studies. 
Our dataset consists of news, rumors, and  hoax which covers a range of different topics, including politics (725 stories), business (224 stories), health (192 stories), and crime (141 stories).

\subsection{Interpretable Models}
Following the previous NLP algorithms for fake news detection, We implement an ensemble of four classifiers for fake news detection to generate different types of explanations. 
Our purpose in choosing the ensemble model approach was to study the effects of different explanation types later in the evaluation experiments. 
We obtain the final prediction score through averaged ensemble results with 73.65\% detection accuracy. 
We review details of each model in the following.

Our first model is a Bi-LSTM network~\cite{huang2015bidirectional} with an additional self-attention layer to extract attention scores for instance explanations.
This model is trained on news headlines only and generates attention explanations for its predictions.
In our empirical tests, Bi-LSTM network outperformed similar networks (e.g., RNN, LSTM. RCNN) for our dataset by capturing both forward and backward states.  
We also use Word2Vec~\cite{mikolov2013distributed} to embed each word into an embedding vector before feeding into the network. 
This trained model achieves 72.00\% fake news detection accuracy on our test set.

The second model performs fake news detection based on both the news story headlines and the set of related articles for each.
The article set representation is constructed using hierarchical attention at sentence level and article level. 
We use the hierarchical attention network (HAN)~\cite{yang2016hierarchical} to help our model focus on the salient sentences and articles at two levels. 
HAN scores each article and selects the most important sentences in each article. 
Each sentence representation of an input article is generated by taking an average of the word embedding of all the words therein. 
Our design allows us to get the attribution score for each article and select the three most important sentences in each article using attention weighs.  
For the news story representation, similar to our first model, we used a Bi-LSTM network.
Finally, a weighted sum is performed over all articles to build the article representation, which is combined with the news story representation to form the final vector representation for news story classification. 
This model achieved 76.04\% classification accuracy on the test set. 

For the third model, we use a knowledge distillation approach~\cite{hinton2015distilling} to approximate a deep architecture (teacher) with a random forest (student) model. 
This model takes news stories, related articles, and article source as the input, and with the mimic learning framework, we can leverage the performance of a deep model and analyze the attribute importance of news stories, articles, and their sources for each prediction. 
We first train a Bi-LSTM teacher model using Glove word embedding~\cite{pennington2014glove} and then train a 60 trees XGBoost~\cite{chen2016xgboost} student model.  
The XGBoost student model provides attribute importance (news story headline, article content, and article source) as for instance explanations. 
Our third model achieved \emph{72.08\%} prediction accuracy in the test set.    

For the last model, we use both news headlines and related articles to train a BiLSTM network with Word2Vec word embedding. 
We use an attention mechanism to focus on parts of the articles that are more relevant to the news story. 
In order to do so, we calculate a weighted average of the hidden state representation based on the attention score corresponding to all the article tokens~\cite{kumar2019attentional}.  
Our method then aggregates all the information about the news story, article context, and attention weights to predict the story's credibility. 
Finally, to generate an overall credibility label for the classification task, the final representation is processed using the final fully connected layer. 
The attention mechanism also generates keyword attribution explanations for each article.

\section{Experimental Design}

We design controlled human subject studies in order to test our hypothesis regarding the effectiveness of AI assistance and its explanation in news review task. 
We present our study design details in terms of study conditions, evaluation measures, and participants' task.

\begin{table*}[]
\centering
\caption{Study conditions and intelligent assistant components to detect fake news and explain its prediction.}
\label{tab:exp-table}
\resizebox{\textwidth}{!}{%
\begin{tabular}{@{}ccl@{}}
\toprule
Study Condition & Model Output & \multicolumn{1}{c}{Model Explanations} \\ \midrule
Baseline & -- & \multicolumn{1}{c}{--} \\ \midrule
AI Assistant & Prediction and Confidence & \multicolumn{1}{c}{--} \\ \midrule
\multirow{3}{*}{\begin{tabular}[c]{@{}c@{}}XAI Assistants\\ (3 conditions)\end{tabular}} & \multirow{3}{*}{Prediction and Confidence} & XAI-attention: Keywork importance heatmap for news headline and articles. \\ \cmidrule(l){3-3} 
 &  & XAI-attribution: News attribute and article importance for related articles. \\ \cmidrule(l){3-3} 
 &  & XAI-all: Explanations from both XAI-attribute and XAI-attention conditions. \\ \bottomrule
\end{tabular}%
}
\end{table*}

\subsection{Study Design}

We conducted human-subjects studies for controlled comparison of elements of the AI assistant and its explanations.
The study followed a between-subjects design with five different conditions, where each participant used one variation of the news reviewing system as described in the following and summarized in Table~\ref{tab:exp-table}. 

\subsubsection{Baseline Condition:}
For the \textit{Baseline} condition, we remove AI prediction and its explanations in the interface.
The baseline interface (Figure~\ref{fig:interface}, top) allows the user to review and share news headlines without any machine learning support. 
This condition serves as the baseline for human-alone performance in comparison to human-AI collaboration. 
Also, since the \textit{Baseline} condition did not include AI or XAI elements, the condition did not measure user mental model and trust in AI or XAI.

\subsubsection{AI Assistant:}
Our interface in \textit{AI Assistant} condition includes AI prediction and confidence for news headline credibility. 
The prediction and confidence from the ensemble model (without explanations) are used in this condition.
The AI predictions are in form of on-demand using a collapsible menu on user click. 
This condition serves as the baseline for user mental model and trust measurements in the AI without explanation.
Figure~\ref{fig:interface} shows model prediction and confidence at (E) and models confidence for the headline and articles separately at (F).

\subsubsection{XAI Assistants:}

The user interface in \textit{XAI assistant} conditions provides instance explanations in addition to news credibility prediction. 
We design three \textit{XAI Assistant} conditions to study how different types of explanations affect Human-AI collaboration. 
We use two interpretable models in each \textit{XAI Assistant} condition. 
The \textit{XAI-attention} condition presents a heatmap of keywords using attention weights for news headline (Figure~\ref{fig:interface}-D) and each related news articles.  
The \textit{XAI-attribution} condition shows news attribution explanations for related articles and news sources. 
The hierarchical attention network generates articles importance score (Figure~\ref{fig:interface}-G) and top-3 important sentences from each article.
Our mimic model generates source, article, and news story attribution score (Figure~\ref{fig:interface}-H) to present instance explanations.
The \textit{XAI-all} condition is the combination of explanations in the \textit{XAI-attribution} and \textit{XAI-attention} conditions.
The purpose for designing \textit{XAI-all} condition was to study the effect of variety of explanation types on users.

\subsection{Study Measures} 

We take users' mental model, human-AI performance, and trust as the primary measures in our studies. 
We mainly use quantitative methods for our measurements to aim for investigating the initial research questions (RQ1 -- RQ3).

\subsubsection{Task Performance:} We calculate the veracity of participants' final shared and reported news as the main performance metric. 
We take the credibility score of user shared news as the number of shared true news divided by total shared news (equal to 12 in all experiments).
We also review and analyze results for the incredibility score (calculated as 1.0 -- \textit{credibility score}) of all reported fake news as the secondary performance measures. 

\subsubsection{Mental Model:} We take participants' accuracy in guessing model output (similar to Poursabzi et al.~\shortcite{poursabzi2018manipulating}) as representative for model predictability and user mental model. 
For the measurement of this prediction task, we use short pop-up questions during the study to ask ``what would the AI fake news detector predict for this news story?'' from participants. 
Participants could response with short ``True'' or ``Fake'' answers. 
Since we expect participants to interact and understand the intelligent assistant during the early stages of the study, the pop-up questions for mental model measurements were limited to the final third of the study (i.e., the last four news review instances).


\subsubsection{User Trust and Reliance:} We measure user trust using a subjective rating of participants' perceived accuracy of AI assistant. 
Specifically, participants answer ``What was the accuracy of the AI fake news detection?'' using a continuous slider bar (between 0--100\%) to indicate their perception of AI or XAI assistant's accuracy in the post-study survey. 

We also measure user reliance using participants' agreement rate with AI assistant predictions. 
To quantitatively measure participants' reliance on model predictions, similar to~\cite{yin2019understanding}, we calculate user agreement rate as the number of news stories which the participant inspected and agreed with the model prediction (either true or fake news), divided by total number of shared or reported news stories.


\subsection{Study Procedure}

Participants started the task by accepting the information sheet including our approved IRB number and study contact points information.
Next, participants saw step-by-step task instructions with visual guides for all interface components.
Visual instructions include descriptions for the headline and article attribution explanations from XAI assistant. 
Next, participants answered the pre-study questionnaire including text entry and multiple-choice questions. 
Participants then started the main task by reviewing news stories. 

\subsubsection{User Task:} Participants were prompted to review a queue of news stories and share 12 true news for social media users.
To engage participants to review news articles and their explanations, users had to select at least one article that represents the news headline for each news story they chose to share. 
They could always skip to the next news story (as many times as needed) if they were not familiar with the topic. 
The choice of the sharing task and ability to skip unfamiliar topics (unlike work that assumes participants are familiar with a short curated list of news stories~e.g.,~\cite{horne2019rating,nguyen2018believe}) improves the fake news detection task by allowing participants to interact and examine the AI/XAI assistant rather than focusing on news analysis. 
Participants also had the chance to flag news stories as fake if they found headlines to be fake; however, these were not counted toward the required number of shared stories needed for task completion. 
Also, in contrast with previous work, our interface gives a list of related news articles to provide the context of news stories for users. 
Further, unlike~\cite{nguyen2018believe}, participants did not receive feedback of the ground truth after each instance (i.e., whether the model made a correct or wrong prediction) to simulate a real-world scenario in which users do not have immediate access to the credibility of their daily news. 
During the last four news stories (the last third of the study), participants were asked pop-up questions about the AI assistant's prediction before revealing the model prediction; This was done to collect data to estimate user ability to predict the AI's output.

In the end, participants answered a final questionnaire of Likert-scale and slider questions about the AI assistant followed by four open-ended response forms.

\subsection{Participant Pool}

Our XAI system and study were designed for non-expert end users with little knowledge of AI. 
We recruited remote participants from Amazon Mechanical Turk ``Master'' users with above 90\% acceptance rate. 
To encourage participants to spend enough time on the task, we measured task duration and paid flexible time-based compensations. 
The payment was set to \$10 per hour and each participant could only participate once in the task. 
To further ensure data quality for analysis, we filtered data samples based on collected user engagement measures including task duration, number of clicks, and character counts in the final questionnaire form. 

\section{Experiments and Results} 

We ran five between-subject experiments in different interface conditions for hypothesis testing. 
The study had a total of 220 Amazon Mechanical Turk participants with equal participants in each condition.
We removed data from 19 participants who spent less than 10 minutes or had especially low interaction behavior during the task. 
A total of 122.8 hours of study time was recorded for the remaining 201 participants, who on average spent 32.1 minutes (range = [10.3, 90.6] with SD = 20.3) on news review and selection, and 6.5 minutes answering surveys and reading instructions. 

In all experiments, we used the same pre-processed and curated queue of news stories to eliminate potential confounds of different news inputs on experiment groups. 
The composition of news stories was organized to show one fake news for every true news to present a scenario with equal mix of true and fake news. 
Also, we controlled participants' observed accuracy for the AI assistant by fixing the rate of the model error to eliminate the effects of model accuracy on study results; for more discussion of this approach, see: ~\cite{nourani2019effects,papenmeier2019model}. 
Specifically, the news queue was curated to present one wrong prediction (with equal rate of false positives and false negatives instances) from the AI assistant after every three correct predictions (i.e., true positive and true negative samples), which resulted in a consistent overall 75.00\% observed AI accuracy for all participants (participants were not informed or aware of this pattern).
Further, as a decision point for trade-offs between clarity and faithfulness of explanations, we performed simple post-processing of model explanations to remove very small features scores in keyword heatmap and normalized articles' attributions scores.

For statistical analysis, inferential tests used one-way independent ANOVAs to compare the conditions for each measure. 
In the end, we briefly review participants' qualitative feedback to see if they support our quantitative findings.

\subsection{Human-AI Performance}


To answer our first research question, we review and analyze the user performance measure for participants' news reviewing and sharing. 
We run a between-subject experiment with 40 participants in three primary interface conditions: 1) \textit{Baseline} without any intelligent assistant, 2) Interface with the \textit{AI Assistant}, and 3) Interface with the \textit{XAI-all} Assistant. \\

\noindent \textit{Hypothesis 1: Users can share more true news stories with the help of XAI Assistant.}

We report the credibility score of participants' shared news as our primary performance measure. 
Results show the average credibility score is higher than the original news feed (50\% credibility) in all three groups that indicates the overall ability of participants in news review and their engagement with the task. 
Participants shared news in \textit{XAI assistant} condition had the highest average of 75.05\% (range = [61\%, 92\%] with SD = 10.06\%) credibility and \textit{Baseline} had the least credibility with 68.4\% (range = [46\%, 88\%] with SD = 11.5\%) credibility. 
The data met the assumptions for parametric testing for all groups with validation checks passing for data normality (Shapiro-Wilk) and homogeneity of variance (Levene's) tests. 
A significant effect was observed by an ANOVA test with $F(2, 107) = 3.32$ and $p = 0.04$ for the news credibility measure among all three conditions. 
A post-hoc Tukey test showed borderline significance ($p = 0.050$) between the \textit{XAI assistant} and \textit{Baseline} conditions, with higher news credibility scores for participants in \textit{XAI-all} group compared to \textit{Baseline} group.

We use incredibility score (calculated as \textit{1.0} -- \textit{credibility score}) of all reported fake news as the secondary performance measures.
Similar to credibility scores for shared news, the \textit{XAI} group has the highest average incredibility of reported fake news stories with 73.8\% reporting fake news (range = [53\%, 100\%] with SD = 10.7\%).  
An ANOVA test revealed a significant main effect with $F(2, 107) = 3.78$ and $p = 0.026$. 
A Tukey post-hoc test showed participants in the \textit{XAI assistant} condition had ($p = 0.019$) reported fake news significantly more than the \textit{Baseline} condition, even though reporting fake news was not the user's primary task during the study. 
\\ 

\noindent \textit{Implications of Results:} 
The study results show that the \textit{XAI assistant} improved user performance compared to the \textit{Baseline} interface without any intelligent agent.
However,  model explanations did not significantly improve user performance over the \textit{AI assistant} condition. 
Given the unique design challenges in misinformation detection models, this is a positive indicator that an intelligent agent together with model explanations can potentially improve collaborative human-AI news reviewing.


\subsection{Mental Model} 

Our second experiment is designed to answer RQ2 by studying the effects of model explanations on users' mental model. 
We recruited new participants to ran studies for hypothesis testing through comparison of \textit{AI assistant} condition (as the baseline) with three \textit{XAI assistant} conditions (as treatments) in our interface.  \\ 

\noindent \textit{Hypothesis 2: Different types of explanations have different effects on user understanding of intelligent assistants.}

Our measure for quantitative evaluation of mental model is through user prediction task (user guessing of model output). 
User accuracy in their prediction task was highest ($M = 62.20\%$) in the \textit{XAI-all} group and the worst ($M = 54.65\%$) in the \textit{XAI-attention} group. 
An ANOVA test detected a significant main effect with $F(3,149) = 3.16$ and $p = 0.026$ for participants between all four conditions with intelligent assistant. 
A Tukey post-hoc test yielded a significant difference ($p = 0.017$) between the \textit{XAI-attention} and \textit{XAI-all} groups. 
However, no significant pairwise difference was detected between the \textit{AI} group and any of \textit{XAI} groups. 
Note that average user prediction task accuracy in the \textit{XAI-attention} group was lower than the \textit{AI assistance} group, indicating the negative effect of explanations in participants' ability to predict model output. \\


\noindent \textit{Implications of Results:} 
The results show a significant effect of explanation types on user mental model based on the user-prediction task measure.
However, none of the model explanation conditions improved users' accuracy in  prediction. 
Notably, word level attention map explanations for news headline and articles (in the \textit{XAI-attention} condition) had a negative effect on user mental model, potentially due to lower user satisfaction and engagement with the AI assistant. 
The discrepancy between user prediction task accuracy between the three XAI conditions indicates that not all explanations are informative or meaningful for end users to be able to predict model behavior.

\begin{figure}[t!]
\captionsetup[subfigure]{justification=centering}
\centering
   \begin{subfigure}[b]{0.49\columnwidth}
        \includegraphics[width=1.0\columnwidth]{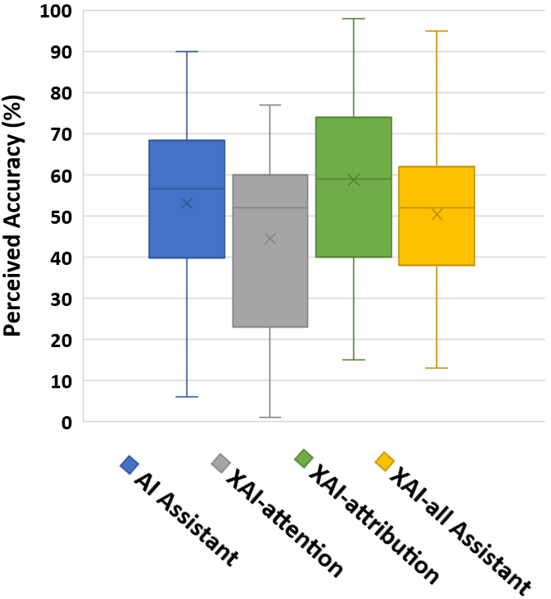}
        \caption{User Perceived Accuracy of Intelligent Assistant}
    \end{subfigure}
      \hfill
    \begin{subfigure}[b]{0.49\columnwidth}
        \includegraphics[width=1.0\columnwidth]{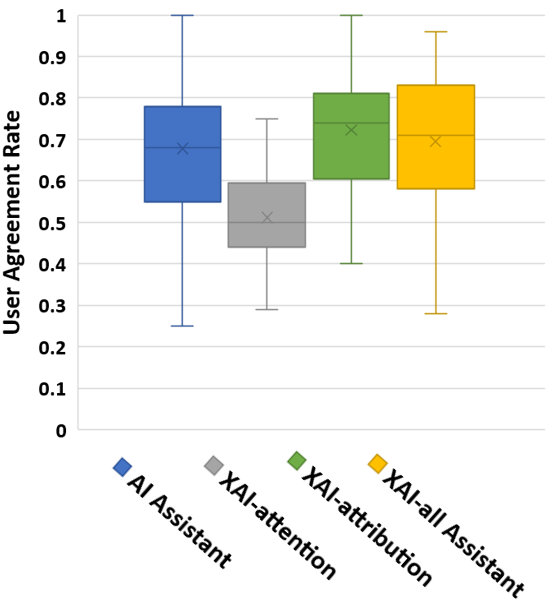}
        \caption{User Agreement Rate with Intelligent Assistant}
    \end{subfigure}
  \caption{User trust measures for \textit{AI Assistant}, and \textit{XAI Assistants} conditions. 
  }
  \label{fig:trust}
\end{figure}

\subsection{Trust and Reliance}

To address RQ3, we review and analyze user trust and reliance measures in our experiments. \\

\noindent \textit{Hypothesis 3: Users have higher perceived accuracy in XAI assistant compared to AI Assistant.}

Our primary measures for user trust in the AI and XAI assistant is the participants' perceived accuracy of the intelligent assistant. 
Figure~\ref{fig:trust}a shows a box-plot of participants' perceived accuracy of the AI and three XAI assistants conditions. 
The results show participants had the highest rate of perceived accuracy in the \textit{XAI-attribution} group (with the visualization of news feature attribution) and lowest in the~\textit{XAI-attention} group (with the heatmap of word feature attribution) on average.
Using an ANOVA test, we found a significant difference ($F(3,155) = 2.86$ and $p = 0.039$) between  perceived accuracy in the four groups. 
For pair analysis, a post-hoc Tukey test revealed participants' perceived model accuracy of the \textit{XAI-attribution} condition ($M = 58.70\%$) was significantly ($p = 0.024$) higher than \textit{XAI-attention} ($M = 45.55\%$). 
Interestingly, participants in the \textit{XAI-all} group responded with lower perceived accuracy ($M = 50.38\%$) compared to AI Assistant ($M = 53.05\%$) with no explanation. \\

\noindent \textit{Hypothesis 4: Users will agree more with XAI assistant predictions compared to AI assistant.} 

We measure user reliance on algorithms via the user agreement rate with AI and XAI assistants predictions. 
Figure~\ref{fig:trust}b presents results for participants agreement rate with the \textit{AI assistant} and three \textit{XAI assistant} groups. 
Overall, participants had near $0.70$ agreement rate with model prediction in all groups except for the \textit{XAI-attention} group with $0.51$ agreement rate. 
We observed a significant main effect using an ANOVA test with $F(3,149) = 16.44$ and $p < 0.001$ for participants agreement rate with intelligent assistants prediction. 
From the pairwise Tukey post-hoc analysis, participants had a significantly lower agreement rate in the \textit{XAI-attention} group compared to all three other groups ($p < 0.001$ for all pairwise comparisons). 
Similar to participants' perceived accuracy, tests did not detect a significant increase in user agreement from model explanations. \\   

\noindent \textit{Implications of Results:}             
The study results indicate that model explanations helped users to adjust their trust and reliance on the intelligent assistant.
We did not observe improvements in user trust or reliance for the XAI assistants over the AI assistant.
In fact, participants actually lost trust in the \textit{XAI-attention} assistant when---despite their initial expectations---they found the system was detecting fake news only based on news keywords.
This could be considered an appropriate result given the limitations of the model logic.
The lower user trust in the \textit{XAI-attention} condition coincides with participants' mental model results and might suggest the effectiveness of explanations in helping users avoid overtrusting the intelligent assistant in cases when model logic may not be optimal or meaningful based on human logic.



\subsection{Qualitative Feedback}

Reviewing participants' written feedback in the post-study survey reflects their reasoning about AI assistant that provides further insight into participants' mental models of the AI/XAI assistants. 
Participants answered two descriptive questions regarding their mental model of the AI assistant's reasoning (``\textit{How do you describe this AI's reasoning to find fake news?}'') and AI assistant's limitations (``\textit{In your opinion, what are the biggest limitations of this AI fake news detector?}''). 
Two authors separately reviewed participants' qualitative feedback and performed open coding to extract themes in participants' notes and comments. 
Two authors coded participants' free response questions to identify salient themes. 
Over three sessions of coding and discussion, they identified 19 codes with an inter-rater reliability of 0.82. 
We use codes to from two main categorizes of responses: AI reasoning, AI limitations, and participant-strategy.

Regarding participant mental models of AI assistants, we observed that explanations clearly improved their understanding of AI reasoning. 
On average, 63.5\% of participants in the \textit{XAI-attention} and 52.8\% of in the \textit{XAI-all} group pointed out the importance of keywords in the news; example comments include ``\textit{I think it looked for certain key words}'' and 
``\textit{The AI compares relevant phrases in the headline to relevant keywords in the supporting stories.}''
In contrast, only 17.9\% of participants in the \textit{AI assistant} condition had expressed such understanding.
We also found 62.0\% participants in the \textit{XAI-attribution} group mentioned related articles and their sources as key features for AI reasoning compared to 31.7\% in the \textit{XAI-attention} group.
For example, one participant in this group commented ``\textit{It tries to pull related articles from the web to prove or disprove the headline}'', and another participant said ``\textit{I think it went by how many article below matched the news.} 

We also found interesting feedback on participants' subjective opinions on the limitations of the AI assistant. 
We saw a clear theme in responses of the need for common sense to distinguish fake and true news. 
On average, from 20\% of participants in all conditions (except \textit{XAI-all} with 11.1\%), we received comments such as ``\textit{it doesn't have human judgment}'', ``\textit{I guess they will not see common sense}'', and ``\textit{The AI doesn't have the experience that a real person has in dealing with the fake news out there.}''. 
Also, participants in \textit{XAI-attention} group paid more attention to the quality and combination of articles in each news story with 43.1\% of them expressing comments like ``\textit{AI doesn't have enough information}'' and ``\textit{It doesn't see multiple sides of the story}'' compared to other conditions with the average of 19.3\%.
Additionally, 27.2\% of all participants expressed concern about AI ability in understanding the context of the news or recognizing sarcasm. 
For instance, one said ``\textit{I think it can't detect sarcasm satire or parody so it has some limitations}'' and another mentioned that ``\textit{The AI isn't able to understand the context of the text. It's not able to actually understand the story or [its] plausibility.}''.

Challenges participants encountered in learning the model behavior was also reflected in 13.5\% of participants' comments in all groups, for example one said: 

\begin{quote}
\textit{I said [to myself] twice that I thought I understood how it worked but when asked to predict the AI's inference about a given headline in the last portion of the study I believe I only matched one out of four so maybe I didn't understand anything that well.}
\end{quote}

Overall, the qualitative user feedback complement the quantitative findings in showing which model explanations helped participants to observe model limitations and adjust their trust and reliance accordingly.






\section{Interactions Between XAI Measures}

\begin{figure}
\captionsetup[subfigure]{justification=centering}
\centering
        \includegraphics[width=1.0\columnwidth]{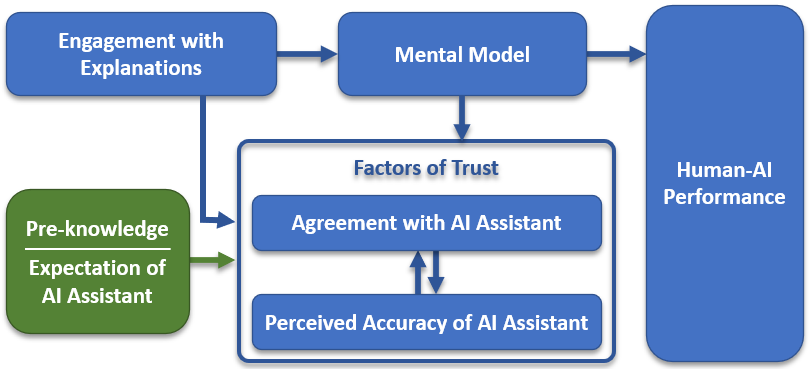}
  \caption{Conceptual model of relationships among user engagement, mental model, trust, and human-AI performance in XAI systems. Figure created based on a model of the ``process of explaining'' in XAI context from~\cite{hoffman2018metrics}. 
  }
  \label{fig:conceptual_model}
\end{figure}

In this section, we summarize different implications of our study results for machine learning explanations and fake news detection. 
Following Hoffman et al.'s~\shortcite{hoffman2018metrics} conceptual model (Figure~\ref{fig:conceptual_model}), we look for correlations between our measurements of user engagement, mental model, performance, and trust to investigate the interplay between these factors.

\subsection{User Expectations of AI Assistant}

We first analyze the relation between user expectations of AI before the study and their perceived algorithm accuracy after the study. 
Research shows that various external and internal factors can interact with user trust, with examples including user pre-knowledge~\cite{horne2019rating}, model stated performance~\cite{yin2019understanding}, and model observed performance~\cite{nourani2019effects}. 
In the pre-study questionnaire, we measured 1) participant expectation of AI assistant accuracy and (with ``If you had an Artificial Intelligence (AI) algorithm to review your daily news for fake news detection, what would be your expectation of AI accuracy to do a good job?'' question) and 2) participant estimation of fake news rate in media (with ``In your experience, what percentage of news that you read daily is false news? e.g., fake news, hoax, rumors, made up stories, misinformation'' question).

As expected, a Pearson test shows a positive correlation ($r = 0.223$, $p = 0.005$) between participants' perceived accuracy at the end of the study and their initial expectation of AI accuracy. 
Regarding participants' expectations of fake news occurrence in daily news, we expected to see more user engagement for participants with higher anticipation of fake news. 
However, we surprisingly found that participants expectation on fake news occurrence has a negative correlation with their engagement with AI assistant ($r = -0.189$, $p = 0.018$).   
This could be due to participants underestimating the AI assistant or choosing their intuition rather than model suggestions. 

\subsection{Engagement with Intelligent Assistants}

As an objective measure of user engagement with intelligent assistants, we consider total continued usage based on the frequency of user interactions (clicks count) with the AI and XAI assistant predictions.
Overall average results show that participants in the \textit{XAI-all} group had the highest engagement rate with the XAI assistant ($0.95$ prediction check rate) for shared or reported news stories. 
An ANOVA test of user engagement with the AI assistant found a significant difference with $F(3,156) =  2.773$ and $p = 0.046$ between conditions, and a Tukey post-hoc test shows participants were significantly ($p = 0.034$) more engaged in the \textit{XAI-all} condition compared to \textit{XAI-attention} condition.

The conceptual model of the process of explaining~\cite{hoffman2018metrics} suggests that explanations in XAI system revise mental model and can engender appropriate trust, see Figure~\ref{fig:conceptual_model}.
To test the interplay between user engagement and their mental model of XAI assistants, we performed a bivariate Pearson correlation test between user engagement rate and prediction task accuracy as the mental model measure. 
Despite the initial hypotheses, a Pearson correlation did not show a positive relation between engagement and mental model ($r = 0.099, p = 0.215$). 
This could be due to the narrow scope of mental-model measurement in our study being limited to the user prediction task (model predictability for users).
However, user engagement had a significant positive correlation ($r = 0.228$, $p < 0.001$) with user agreement with the intelligent assistant. 
This shows as more participants got involved with the AI or XAI predictions, the more they agreed with its predictions.

\subsection{Mental Model Affecting Performance and Trust} 



Next, we analyze how users' mental model interacts with trust and human-AI performance. 
A Pearson test between users' prediction task accuracy (mental model measure) and perceived accuracy of AI assistant (our first user trust measure) showed a positive significant correlation ($r = 0.212$, $p = 0.008$). 
A correlation test between user prediction accuracy and user agreement with the AI assistant (our second user-trust measure) also showed a positive significant correlation ($r = 0.280$, $p < 0.001$) between participants' mental model and trust.
Positive correlations of mental model with both trust measures demonstrate the relation between predictability of the intelligent agent and trust.

As hypothesized, user prediction task accuracy was positively correlated with credibility of shared news ($r = 0.305$, $p < 0.001$) as well as incredibility of reported fake news ($r = 0.283$, $p < 0.001$).
This finding suggests users with a more accurate mental model could better guess model failure cases, and by avoiding those cases, they could improve their performance.

\subsection{Interactions Between Trust Measures}

Another interesting finding from our study is that we observed interactions between multiple measures of user trust.
Previous research studies have utilized various independent trust measures such as perceived algorithm performance~\cite{nourani2019effects}, perception of control over the system~\cite{Kocielnik2019will}, and the rate of user agreement with an algorithm's recommendations~\cite{yin2019understanding}.  
In our studies, we measured two different trust factors to examine how they may interact. 
A Pearson correlation test between the two trust measures shows a positive significant correlation between the perceived accuracy and user agreement rate ($r = 0.482$, $p < 0.001$). 
This positive correlation suggests that as users feel more confident about AI competence, they tend to agree more with its predictions.





\section{Conclusion and Future Work} 

In our research, we evaluated model explanations from multiple models as part of an ensemble approach for fake news detection. 
This approach allowed us to study how different types of explanations affect users in fake news detection. 

In conclusion, our research revealed multiple challenges in designing effective XAI systems in the fake news detection domain. 
In particular, we observe challenges rising from the inherent difference between models' feature learning (word-level features in our case) and human understanding of news and information. 
Overall, users' interaction with the AI and XAI assistants affected their performance, mental model, and trust. 
However, model explanations in our studies did not improve task performance or increase user trust and  mental model. 
Instead, the quantitative results and qualitative feedback indicate that explanations helped users' to build an appropriate mental model of intelligent assistants and adjust their trust accordingly given the limitations of the models. 
For example, participants in the \textit{XAI-attention} group that was significantly less successful in guessing model outputs also showed significantly lower trust (in both trust measures) compared to the \textit{XAI-all} condition. 
Likewise, reviewing user engagement results showed that~\textit{XAI-attention} explanations were not appreciated by the users. 
Therefore, we conclude that improving transparency of the model helped users to appropriately avoid overtrusting the fake news detector when they found the AI reasoning was not trustworthy or simply explanations were nonsensical. 
Future research is needed to assess the effectiveness of other types of explanations, such as knowledge graphs and multi-modal evidence retrieval on users in fake news detection assistants.




\bibliographystyle{aaai}
\bibliography{biblo}

\end{document}